\documentstyle[sprocl]{article}
\begin{document}
\title{LEPTON $CP$ ASYMMETRY IN $B$ FACTORIES: \\
$~\epsilon^{~}_B$ AND NEW PHYSICS} 
\author{\bf Zhi-zhong Xing}
\address{
Physics Department, Nagoya University, Chikusa-ku,
Nagoya 464-01, Japan 
}
\maketitle
\abstracts{
In this talk, we first make some brief comments on the phase-convention 
dependence of the $CP$-violating parameter $\epsilon^{~}_B$.
A simple framework is then presented for analyzing new physics contributions
to $B^0$-$\bar{B}^0$ mixing and their effects on the lepton $CP$
asymmetry, a promising signal to be further searched for in the first-round
experiments of $B$ factories.
}

\section{Introduction}

The study of various possible $CP$-violating phenomena in weak $B$-meson
decays, starting from the end of 1970's, becomes more intensive
today -- on the eve of KEK and SLAC $B$ factories, which will provide
a unique opportunity to test the Kobayashi-Maskawa (KM) mechanism of $CP$
violation in the standard model (SM)
and to discover possible new physics (NP) beyond the SM.

Phenomenologically three different types of $CP$-violating
signals are expected in neutral $B$-meson transitions:
(a) $CP$ violation from $B^0$-$\bar{B}^0$ mixing; (b) 
$CP$ violation from direct $b$-quark decay; and (c) $CP$ violation
from the interplay of decay and mixing. Type (a) can 
be detected most appropriately from the decay-rate
asymmetry between two semileptonic channels $B^0 \rightarrow l^+
\nu^{~}_l X^-$ and $\bar{B}^0 \rightarrow l^- \bar{\nu}^{~}_l X^+$,
the so-called ``lepton $CP$ asymmetry'' (denoted as ${\cal A}_{\rm
SL}$ below). Type (b) may involve large
uncertainties associated with the evaluation of hadronic matrix
elements and (or) penguin contributions. Type (c) should play a
crucial role in determining the KM $CP$-violating phase and (or) the
NP phase(s), as it is independent of hadron or penguin pollution to 
a good degree of accuracy in some decay modes \cite{Sanda}. 

Experimentally, however, only the lepton $CP$ asymmetry ${\cal A}_{\rm 
SL}$ has been searched for to date. An upper bound on this asymmetry 
was first set by the CLEO time-integrated measurement: 
$|{\cal A}_{\rm SL}| < 0.18$ or ${\rm Re}\epsilon^{~}_B < 0.045$
at the $90\%$ confidence level \cite{CLEO,PDG96}, where $\epsilon^{~}_B$ is
a $CP$-violating parameter of $B^0$-$\bar{B}^0$ mixing defined like
$\epsilon^{~}_K$ in the neutral kaon system.  
The OPAL time-dependent measurement has recently 
given a more accurate result \cite{OPAL}:
${\rm Re}\epsilon^{~}_B = 0.002 \pm 0.007 \pm 0.003$ in a special
convention used for $\epsilon^{~}_B$, which is
equivalent to ${\cal A}_{\rm SL} = 0.008 \pm 0.028 \pm 0.012$.
Both experiments are consistent with the 
standard model prediction, i.e., $|{\cal A}_{\rm SL}| \sim 10^{-3}$
or smaller. It should be noted, nevertheless, 
the existence of NP in $B^0$-$\bar{B}^0$
mixing is possible to enhance ${\cal A}_{\rm SL}$ up to the percent
level \cite{SX97}, observable in the first-round experiments
of a $B$-meson factory with about $10^{8}$ events of $B^0\bar{B}^0$
mesons \cite{Yamamoto}. 

The purposes of this talk are: (1) to comment on the phase convention
dependence of the parameter $\epsilon^{~}_B$; and (2) to present a
simple framework for analyzing NP effect on ${\cal A}_{\rm SL}$. 
We expect that the further measurement of ${\cal A}_{\rm SL}$ and its
correlation with other $CP$ asymmetries may serve as a sensitive probe 
for NP in $B^0$-$\bar{B}^0$ mixing.

\section{Comments on $\epsilon^{~}_B$}

In some literature, the $CP$-violating parameter $\epsilon^{~}_B$ is
defined to relate the flavor eigenstates $|B^0\rangle$ and
$|\bar{B}^0\rangle$ to the mass eigenstates $|B_1\rangle$ and
$|B_2\rangle$:
\begin{equation}
|B_{1,2} \rangle \; = \; \frac{1}{\sqrt{2(1+|\epsilon^{~}_B|^2)}}
\left [ \left ( 1 + \epsilon^{~}_B \right ) |B^0\rangle ~ \pm ~ 
\left ( 1 - \epsilon^{~}_B \right ) |\bar{B}^0\rangle \right ] \; .
\end{equation}
Certainly $\epsilon^{~}_B$ has the phase freedom arising from the
bound states $|B^0\rangle$ and $|\bar{B}^0\rangle$. The reason is simply
that $|B^0\rangle$ and $|\bar{B}^0\rangle$ are
defined by strong interactions only, leaving the relative phase
between them undetermined. The dependence of $\epsilon^{~}_B$ on the
bound-state phase convention can be removed, if one adopts the
definition to link the $CP$ eigenstates $|B_+\rangle$ ($CP$-even) and
$|B_-\rangle$ ($CP$-odd) with the mass eigenstates $|B_1\rangle$ and
$|B_2\rangle$: 
\begin{equation}
|B_{1,2} \rangle \; = \; \frac{1}{\sqrt{1 + |\varepsilon^{~}_B|^2}} \left (
|B_{\pm}\rangle ~ + ~ \varepsilon^{~}_B |B_{\mp}\rangle \right ) \; .
\end{equation}
Note that $\varepsilon^{~}_B$ and $\epsilon^{~}_B$ are not identical \cite{BB},
unless we fix the $CP$ transformation between $|B^0\rangle$ and
$|\bar{B}^0\rangle$ as $(CP)|B^0\rangle = \pm |\bar{B}^0\rangle$.
Therefore it is better to
adopt the phase-convention independent parameter $\varepsilon^{~}_B$, other than 
$\epsilon^{~}_B$, in phenomenological applications \cite{Tsai}. In the analyses of 
CLEO and OPAL measurements \cite{CLEO,OPAL}, however, the
phase-convention dependent parameter $\epsilon^{~}_B$ has been used.

The more important point is that both $\varepsilon^{~}_B$ and
$\epsilon^{~}_B$ depend on the phase convention of the KM matrix for
quark flavor mixings, although $\varepsilon^{~}_B$ itself is
independent of the bound-state phases of $|B^0\rangle$ and
$|\bar{B}^0\rangle$. The KM phase convention comes from the freedom in 
defining the quark field phases, thus different parametrizations of
the KM matrix may result in different values for $\varepsilon^{~}_B$
and $\epsilon^{~}_B$. In this sense, even $\varepsilon^{~}_B$ is {\it not} a 
physically well-defined parameter. 

In the CLEO and OPAL analyses \cite{CLEO,OPAL}, ${\rm
Im}\epsilon^{~}_B =0$ or $|{\rm Im}\epsilon^{~}_B| \ll 1$ has been
assumed in an undemonstrative way \cite{Nir,Xing}. 
That is why these analyses can yield an 
upper bound on ${\rm Re}\epsilon^{~}_B$ solely from the 
measurement of the lepton $CP$
asymmetry. To see this point more clearly, let us take the CLEO
constraint on ${\rm Re}\epsilon^{~}_B$ into account. The dilepton $CP$ 
asymmetry measured on the $\Upsilon (4S)$ resonance reads
\begin{equation}
{\cal A}_{\rm SL} (\varepsilon^{~}_B) \; =\; 
\frac{4 {\rm Re}\varepsilon^{~}_B \left ( 1 + |\varepsilon^{~}_B|^2
\right )}{ \left (1 + |\varepsilon^{~}_B|^2 \right )^2 + 4 ({\rm
Re} \varepsilon^{~}_B )^2} \; ,
\end{equation}
and ${\cal A}_{\rm SL}(\epsilon^{~}_B) = {\cal A}_{\rm
SL}(\varepsilon^{~}_B)$ holds. Only in the assumption $|{\rm
Im}\epsilon^{~}_B| \ll 1$ (and $|{\rm Re}\epsilon^{~}_B|\ll 1$), 
one can get ${\cal A}_{\rm SL}
(\epsilon^{~}_B) \approx 4 {\rm Re} \epsilon^{~}_B$. Then $|{\cal
A}_{\rm SL} (\epsilon^{~}_B)| < 0.18$ leads to 
$|{\rm Re}\epsilon^{~}_B| < 0.045$. If
one takes $|{\rm Im}\epsilon^{~}_B| \sim 1$, however, 
${\cal A}_{\rm SL}(\epsilon^{~}_B) \approx 2 {\rm Re}\epsilon^{~}_B$
appears \cite{Pakvasa} and a different upper bound on ${\rm Re}\epsilon^{~}_B$ 
must turn out from the same measurement of ${\cal A}_{\rm SL}
(\epsilon^{~}_B)$. Indeed the condition ${\rm Im}\epsilon^{~}_B = 0$ 
corresponds to a special parametrization of the KM matrix which has
${\rm Im}V_{td} = {\rm Im}V_{tb} =0$ (e.g., the one proposed recently
by Fritzsch and the author \cite{FX97}). In contrast, the condition 
$|{\rm Im}\epsilon^{~}_B| \sim O(1)$ may be satisfied in the
``standard'' parametrization or the Wolfenstein form \cite{Xing}.

\section{NP effect on ${\cal A}_{\rm SL}$}

Now we present a simple framework to analyze NP contributions to
${\cal A}_{\rm SL}$ and other $CP$ asymmetries via $B^0$-$\bar{B}^0$
mixing. In terms of the off-diagonal elements of the $2\times 2$
$B^0$-$\bar{B}^0$ mixing Hamiltonian $\bf M - {\it i} \Gamma/2$,
${\cal A}_{\rm SL}$ can be written as
\begin{equation}
{\cal A}_{\rm SL} \; = \; {\rm Im} \left ( \frac{\Gamma_{12}}{M_{12}} 
\right ) \; .
\end{equation}
In most extensions of the SM, NP can significantly
contribute to $M_{12}$. However, NP is not expected to significantly affect 
the direct $B$-meson decays via the tree-level $W$-mediated channels.
Thus $\Gamma_{12} = \Gamma^{\rm SM}_{12}$ holds
as a good approximation, where \cite{SX97}
\begin{equation}
\Gamma^{\rm SM}_{12} \; =\; - ~ \frac{G^2_{\rm F} ~ B_B ~ f^2_B ~ M_B
~ m^2_b}{8\pi} ~ \left [ (\xi^*_u)^2 T_u + (\xi^*_c)^2
T_c + (\xi^*_t)^2 T_t \right ] \; 
\end{equation}
with KM factors $\xi_i \equiv V^*_{ib} V_{id}$ ($i=u,c,t$) for three quark
families and QCD correction factors $T_u \sim - T_c \sim 0.1$ and $T_t 
\sim 1$. In the presence of NP, $M_{12}$ can be written as
\begin{equation}
M_{12} \; =\; M^{\rm SM}_{12} ~ + ~ M^{\rm NP}_{12} \; .
\end{equation}
Rescaling three complex quantities in Eq. (6) by
$|M_{12}| = \Delta M /2$, where $\Delta M = (0.464\pm 0.018) ~ {\rm
ps}^{-1}$ has been measured \cite{PDG96}, we obtain a parametrized 
triangle in the complex plane (see Fig. 1).
\begin{figure}[t]
\begin{picture}(400,160)(-35,210)
\multiput(60,300)(4,0){40}{\line(1,0){2}}

\put(80,300){\line(1,3){22}}
\put(80,300.5){\line(1,3){22}}
\put(80,299.5){\line(1,3){22}}
\put(80,300){\vector(1,3){11}}
\put(80,300.5){\vector(1,3){11}}
\put(80,299.5){\vector(1,3){11}}
\put(81,336){\makebox(0,0){1}}

\put(102,366){\line(3,-2){47}}
\put(102,366.5){\line(3,-2){47}}
\put(149,334.5){\vector(-3,2){24}}
\put(149,335){\vector(-3,2){24}}
\put(140,355){\makebox(0,0){\scriptsize $R_{\rm NP}$}}

\multiput(149,334.5)(3,-2){18}{\circle*{0.3}}

\put(80,300){\line(2,1){69}}
\put(80,300.5){\line(2,1){69}}
\put(80,300){\vector(2,1){35}}
\put(80,300){\vector(2,1){35}}
\put(113,330){\makebox(0,0){\scriptsize $R_{\rm SM}$}}

\put(119,308){\makebox(0,0){\scriptsize $2\phi_1$}}
\put(203,308){\makebox(0,0){\scriptsize $2\theta$}}
\end{picture}
\vspace{-3.5cm}
\caption{\small Triangular relation of $M^{\rm SM}_{12}$, $M^{\rm
NP}_{12}$ and $M_{12}$ (rescaled by $\Delta M/2$).}
\end{figure}
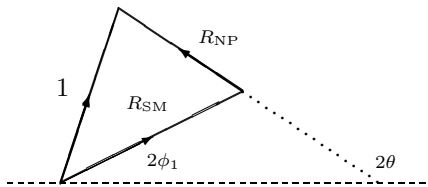
The correlation between the SM and NP parameters reads
\begin{equation}
R_{\rm NP} \; =\; - R_{\rm SM} ~ \cos 2 (\theta - \phi_1) ~
\pm ~ \sqrt{1 - R^2_{\rm SM} ~ \sin^2 2 (\theta - \phi_1)} \; ,
\end{equation}
where $R_{\rm SM}$ can be calculated in the SM box-diagram
approximation \cite{SX97}:
\begin{equation}
R_{\rm SM} \; = \; \frac{G^2_{\rm F} ~ B_B ~ f^2_B ~ M_B ~ m^2_t}{6\pi^2 ~
\Delta M} ~ \eta^{~}_B ~ F\left (\frac{m^2_t}{m^2_W}\right ) ~ |\xi_t|^2 \; 
\end{equation}
with $\eta_B \approx F(m^2_t/m^2_W) \approx 0.55$.
We see that there exist two solutions for $R_{\rm NP}$.

The lepton $CP$ asymmetry ${\cal A}_{\rm SL}$ turns out to be \cite{SX97}:
\small
\begin{eqnarray}
\frac{{\cal A}_{\rm SL}}{C_m} & = & R_{\rm SM} R_{\rm NP} \left [ {\rm Im} 
\left ( \frac{\xi_u}{|\xi_t|}\right 
)^2 T_u  + {\rm Im} \left ( \frac{\xi_c}{|\xi_t|}\right )^2 T_c
+ {\rm Im} \left ( \frac{\xi_t}{|\xi_t|}\right )^2
T_t \right ] \cos (2\theta) \nonumber \\
&  & + ~ R^2_{\rm SM} \left [ {\rm Im} \left
( \frac{\xi_u}{\xi_t} \right )^2 T_u + {\rm Im} \left
( \frac{\xi_c}{\xi_t} \right )^2 T_c \right ] \; + \; \\ 
&  & R_{\rm SM} R_{\rm NP} \left [ {\rm Re} 
\left ( \frac{\xi_u}{|\xi_t|}\right 
)^2 T_u + {\rm Re} \left ( \frac{\xi_c}{|\xi_t|}\right )^2 T_c
+ {\rm Re} \left ( \frac{\xi_t}{|\xi_t|}\right )^2
T_t \right ] \sin (2\theta) \; , \nonumber 
\end{eqnarray}
\normalsize
where $C_m \approx 1.3 \times 10^{-2}$. 
Clearly the second term of ${\cal A}_{\rm SL}$ comes purely from
$M^{\rm SM}_{12}$ itself and its magnitude is expected to be of
$O(10^{-3})$ due to the absence of the $T_t$ contribution.
The first and third terms of ${\cal A}_{\rm SL}$ arise from the
interference between $M^{\rm SM}_{12}$ and $M^{\rm NP}_{12}$;
but they depend on nonvanishing ${\rm Im} (M^{\rm SM}_{12})$ and
${\rm Im} (M^{\rm NP}_{12})$, respectively. 
For appropriate values of $\theta$ and $\phi_1$,
magnitudes of both the first and third terms of ${\cal A}_{\rm SL}$ may
be at the percent level! To obtain $|{\cal A}_{\rm SL}| \sim O(10^{-2})$,
however, there should not be large cancellation 
between two dominant terms in Eq. (9).

The $CP$ asymmetry in $B_d \rightarrow J/\psi K_S$ can be calculated in 
the same framework. We obtain
\begin{equation}
{\cal A}_{\psi K} \; =\; R_{\rm SM} \sin
(2\phi_1) ~ + ~ R_{\rm NP} \sin (2\theta)  \; .
\end{equation}
As $R_{\rm NP}$, $R_{\rm SM}$ and $\phi_1$, $\theta$ are dependent
on one another through Eq. (7), $|{\cal A}_{\psi K}| \leq 1$
is always guaranteed within the allowed parameter space.

Two interesting cases, corresponding to ${\rm Im}M^{\rm
SM}_{12} = 0$, ${\rm Im}M^{\rm NP}_{12} \neq 0$ and ${\rm
Im}M^{\rm NP}_{12} = 0$, ${\rm Im}M^{\rm SM}_{12} \neq 0$, have been
numerically illustrated by Sanda and the author \cite{SX97}. It is
found that ${\cal A}_{\rm SL}$ does have good chances to reach the
percent level, and the correlation between ${\cal A}_{\rm SL}$ and
${\cal A}_{\psi K}$ (as well as other $CP$ asymmetries) does reflect
the information from NP.

\section{Summary}

We have commented on the dependence of 
$\epsilon^{~}_B$ or $\varepsilon^{~}_B$ on the KM phase convention. 
We emphasize that neither of them can prove much advantage in
describing data of $CP$ violation from $B^0$-$\bar{B}^0$ mixing.
Also a simple framework has been presented for the analysis of
possible NP contributions to $B^0$-$\bar{B}^0$ mixing and of their
effects on ${\cal A}_{\rm SL}$ and other $CP$ asymmetries. If we are
lucky, we should be able to detect the lepton $CP$ asymmetry at the
percent level in the first-round experiments of $B$ factories.

{\it Acknowledgments:} 
I would like to thank Y.Y. Keum for his warm hospitality
and APCTP for its generous support, which made my participation
in this wonderful workshop realizable. I am also grateful to A.I. Sanda
for sharing physical ideas with me in a recent paper \cite{SX97}, whose
results were partially presented here. Finally useful discussions with
S. Pakvasa, A.I. Sanda, S.Y. Tsai and D.D. Wu are acknowledged.


\begin{thebibliography}{99}
\bibitem{Nir} I would like to thank Y. Nir for pointing out this fact to me.
\end{thebibliography}
\end{document}